\documentclass[twocolumn,superscriptaddress, preprintnumbers,prb]{revtex4-1}
\usepackage{graphicx}
\usepackage{dcolumn}
\usepackage{amssymb}
\usepackage{amsmath}
\usepackage{float}
\usepackage{setspace}
\usepackage{hyperref}
\usepackage{url}
\usepackage{lipsum}

\hypersetup
{
colorlinks=true, 
citecolor=blue
}

\begin{document}

\title{Disorder effects in two-dimensional flat-band system with next-nearest-neighbor hopping}

\author{Yue Heng Liu}
\affiliation{GBA Branch of Aerospace Information Research Institute, Chinese Academy of Sciences, Guangzhou 510535, China}
\affiliation{Guangdong Provincial Key Laboratory of Terahertz Quantum Electromagnetics, Guangzhou 510700, China}

\author{Zi-Xiang Hu}
\affiliation{Department of Physics and Chongqing Key Laboratory for Strongly Coupled Physics, Chongqing University, Chongqing 401331, People's Republic of China}

\author{Qi Li}
\email{liqi@aircas.ac.cn}
\affiliation{GBA Branch of Aerospace Information Research Institute, Chinese Academy of Sciences, Guangzhou 510535, China}
\affiliation{Guangdong Provincial Key Laboratory of Terahertz Quantum Electromagnetics, Guangzhou 510700, China}

\begin{abstract}
For two-dimensional Lieb lattice, while intrinsic spin-orbit coupling is responsible for opening the gap that exhibits the quantum spin Hall effect, topological phase transitions are driven by a real next-nearest-neighbor (NNN) hopping. In this work, we utilize the transfer matrix method to study the flat-band localization mechanism in the presence of complex NNN hoppings.  We demonstrate that the geometric localization in flat bands can be alleviated by topological edge states under weak disorder. Furthermore, correlated disorders are shown to induce inverse Anderson transition with the topological edge  states persisting under strong disorder, a robustness confirmed by Chern number calculations, which identifies the root cause of this phenomenon. These findings establish a unified platform for investigating topological phase transitions, flat bands, and disorder effects.
\end{abstract}

\date{\today}

\maketitle

\section{Introduction}

The study of flat-bands can be traced back to 1986,  when Sutherland first predicted electron localization in flat-bands induced by local topology in Dice model~\cite{Sutherland}.
Flat-bands with fixed and dispersionless eigenenergies imply that the group velocity of a wave packet is zero or the effective mass of particles is infinite. The macroscopic degeneracy allows for the construction of localized flat-bands modes through the superposition of Bloch waves and  nonzero amplitude are only at a finite number of lattice sites.  The most remarkable characteristic of flat-bands is the frozen electrons motion which could give rise to a rich array of strongly correlated physical phenomena including unconventional  superconductivity~\cite{Cao, Balents, Peri},  ferromagnetism~\cite{Mielke1, Mielke2, Mielke3}, fractional quantum Hall effect~\cite{Wang2011, Wang2012, DNSheng} and fractional Chern insulators~\cite{Regnault, Bergholtz, Behrmann, Liu} . 

Based on destructive interference, flat-bands are compact localized states (CLSs) formed with local network symmetry (symmetry-protected ), fine tuned  tight-binding parameters (accidental) or lattice topological structure(topological-protected). Flach's group classifies flat-bands modes according to the number of unit cells $U$ occupied by  compact localized states~\cite{FlachEPL}. Typically,  symmetry protected flat-bands possess a complete set of orthogonal basis states, corresponding to $U=1$. Examples include  one-dimensional (1D) cross-stitch and  rhombic lattice. Modes wtih  $U \geq 2$  correspond to the case of  accidental flat-bands. Their band structures are not always flat and they emerge only incidentally under specific prarmeters~\cite{Mielke1, Tasaki}.  One typical example is the two-dimensional (2D) kagome lattice which is a line graph of a honeycomb lattice. The topological-protected flat-bands are more robust which are determined by the topological structure of the lattice and could remain largely intact even if a band gap opens under certain perturbations. The topological-protected 2D Lieb model is a typical bipartite lattice and possesses chiral symmetry when take only nearest neighbor hoppings~\cite{Lieb}. This model is built on a line-centered square lattice with three sites (A, B, and C) per unit cell. It exhibits a three-band structure featuring one flat-bands that touches the other two dispersive bands to form a Dirac cone at the $\Gamma$ point. Experimentally,  the Lieb lattice can be realized in photonic crystals ~\cite{Vicencio, Silva, Sebabrata, Xia}, optical lattices ~\cite{Shen, Apaja} and cold atom systems~\cite{Goldman}. The 2D Lieb lattice is a topological insulator when considering intrisic spin-orbital coupling (ISOC)~\cite{Weeks1, Goldman}. Topological phase transitions driven by ISOC and real next-nearest-neighbor coupling have been investigated in a lot of research~\cite{Beugeling, Zhao, Weeks2, Chen2017}.  

Although the strongly correlated many-body physics in flat-bands systems is a recent research hotspot, the issues of localization and phase transitions are equally important. It is well established that disorder and impurities, which are ubiquitous in real quantum materials, play a decisive part in bringing about localization and quantum phase transitions. 

For example, with sufficiently strong uncorrelated disorders in lattice, a sudden breakdown of transport and complete localization of wave functions is understood as a result of Anderson localization~\cite{Anderson}. In flat-bands systems, localization and Anderson  transition in two and three-dimensional extended Lieb lattice have been investigated systematically~\cite{Mao, JieLiu2020}.  In contrast to ordinary Anderson localization, under correlated disorders unconventional delocalization or inverse Anderson transition are studied in various lattice models with flat-bands~\cite{Izrailev, FlachPRL, GneitingPRB, JieLiu2022, Zuo, Qi}. In experiment, observation of inverse Anderson localization is achieved via a photonic Aharonov-Bohm cage with the introduction of correlated disorder~\cite{Vidal, Mukherjee, Gligoric, Kremer, Longhi, Yanbo}.
Even more interestingly, theoretical prediction and experimental confirmation tell that strong disorder can not only destroy topological phases, but in some cases, can even induce topological Anderson insulators~\cite{ShenSQ, GrothPRL, Meier, Stutzer, LiuGG, Cui}.

Issues such as disorder, localization, and topological phase transition in flat-band system are of significant research interest. In this paper, we numerically study the interplay of localization, disorder and phase transition in flat-bands systems. Explicitly, we construct the transfer matrix of a 2D Lieb lattice in the presence of complex next-nearest-neighbor couplings. Using the transfer matrix method, we specifically discuss the localization behavior of flat-band under the influence of disorders in different topological phases. Secondly, by introducing correlated disorders, we find that edge states remain stable even within large disorder strengths confirmed by calculating the real-space Chern number via the coupling matrix method. We conjecture that correlated disorder configuration and topological protected edge states contribute to the occurrence of inverse Anderson localization.

\section{Model and Method}
\subsection{2D Lieb lattice with complex NNN hopping}

\begin{figure}[htbp]
\center
\includegraphics[width=9cm]{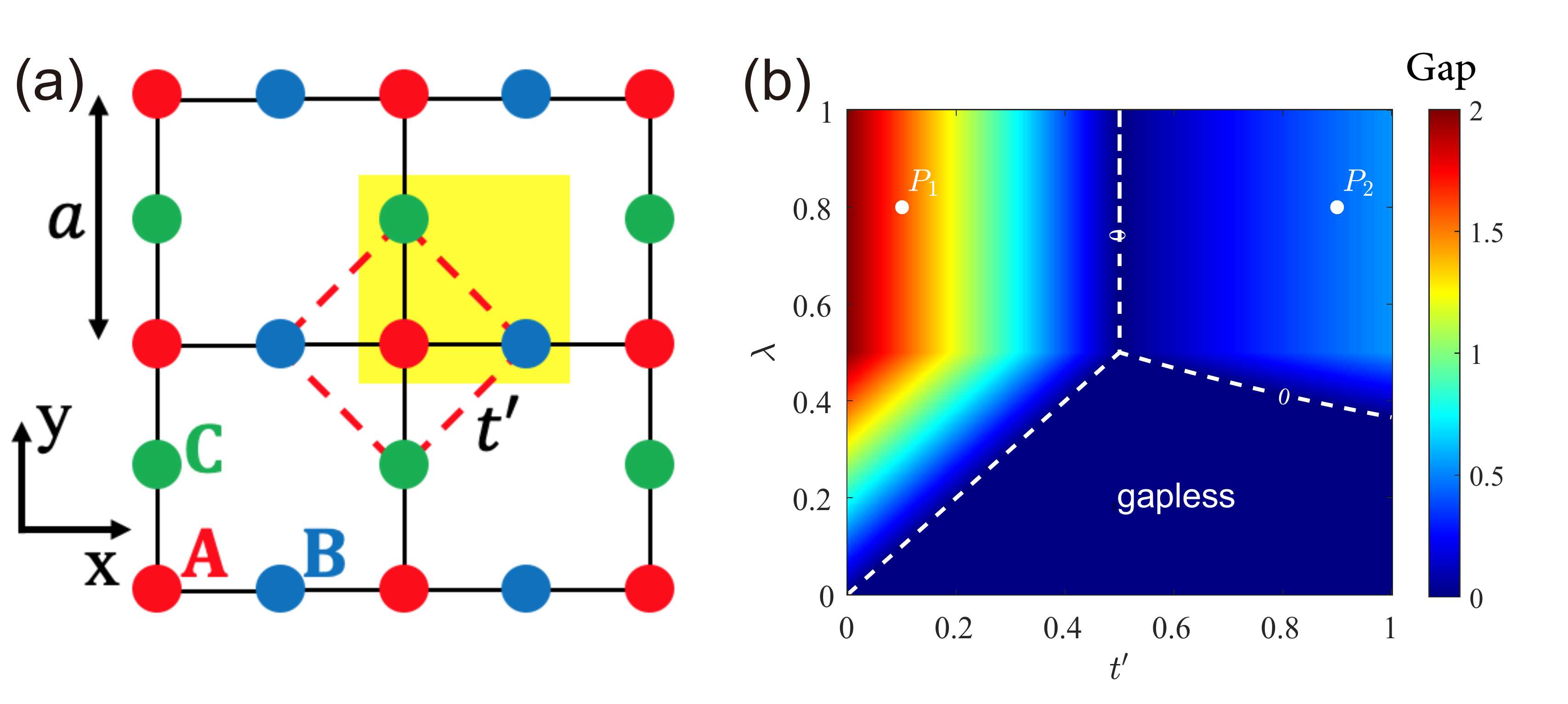}
\caption{(a)2D Lieb lattice. One unit cell contains three sublattice $\{A,B,C\}$ (yellow area). The location of unit cell is labeled by $(m, n)$ in real space. The NN hopping $t$ is set to 1 for simplicity and the NNN hopping between B and C sublattices $t'$ is indicated using dashed red line. The square unit cell has lattice constant $a \equiv 1$. (b) Energy gap (gap2) between top and middle branches as a function of parameter $t'$ and $\lambda$. The two sets of parameters are marked as $P_1(t'=0.1, \lambda=0.8)$ and $P_2(t'=0.9, \lambda=0.8)$. }
\label{fig:sketch-gap}
\end{figure}

First, we investigate  the 2D Lieb lattice (depicted in Fig.\ref{fig:sketch-gap}(a) ),  with three sites A, B and C per unit cell (yellow area). In continuum limit, the spectrum consists of three energy bands, with one flat band $ E_f  = 0 $ and two dispersive bands $E_d(k) = \pm 2t \sqrt{ (\cos \frac{k_x}{2})^2 + (\cos \frac{k_y}{2})^2 }$ where onsite potentials are zero and $t$ denotes the nearest-neighbor coupling between A and B or C sites.  

It is numerically convenient to describe the 2D Lieb lattice structure using the site basis $\Psi_{m,n}=(\Psi_{m,n}^A,\Psi_{m,n}^B,\Psi_{m,n}^C)^T$, where $m$ and $n$ represent the lattice indices in the $x$ and $y$ directions, respectively. In torus geometry with periodic boundary conditions in both  x- and y-directions, the corresponding 2D Lieb tight-binding model can be written as 
\begin{eqnarray}
\label{eq:tb}
E \psi_{m,n}^A &=&  t ( \psi_{m,n}^B+\psi_{m,n}^C +\psi_{m-1,n}^B +\psi_{m,n-1}^C )   \nonumber   \\
&& \delta( \psi_{m-1,n}^A+\psi_{m+1,n}^A+\psi_{m,n-1}^A+\psi_{m,n+1}^A)     \nonumber     \\
E \psi_{m,n}^B &=& t ( \psi_{m,n}^A+\psi_{m+1,n}^A )   \nonumber   \\
&+& (i\lambda-t')( \psi_{m,n-1}^C +\psi_{m+1,n}^C )   \nonumber   \\
&-&  (i\lambda+t') (  \psi_{m,n}^C +\psi_{m+1,n-1}^C )   \nonumber  \\
E \psi_{m,n}^C &=& t ( \psi_{m,n}^A+\psi_{m,n+1}^A )  \nonumber   \\
&+&  (i\lambda-t')( \psi_{m,n}^B +\psi_{m-1,n+1}^B )   \nonumber   \\
&-&  (i\lambda+t') (  \psi_{m,n+1}^B +\psi_{m-1,n}^B )  
\end{eqnarray}

Here we  include  intrinsic spin-orbit coupling with a complex hopping integral  $\lambda  \vec{v}_{ij}$ and unit vector $\vec{v}_{ij}=(\vec{d}_{ik}\times \vec{d}_{kj} ) / \vert \vec{d}_{ik}\times \vec{d}_{kj}  \vert $.  $i$ and $j$ are two next nearest neighbor sites and $k$ their common nearest neighbor. The vector $\vec{d}_{ik}$ points from $i$ to $k$.  The next-nearest-neighbor hopping integral (real number) between sites B and C is denoted by $t'$.
It should be noted that  we introduce hopping $\delta = 0.0001$ between different $A$ sites in adjacent unit cells to facilitate  subsequent calculation of localization length and avoid matrix inversion issues.

For simplicity, we restrict our discussion to a single spin component, noting that the resulting bands are doubly degenerate. Therefore, when specifically considering a Hamiltonian that includes only the spin-up component, we obtain the momentum-space representation of the Hamiltonian through a Fourier transform:

\begin{equation}
H_\textbf{k} = \left(
\begin{matrix}
2\delta( \cos k_x + \cos k_y) & -2t \cos \frac{k_x}{2} &   -2t \cos \frac{k_y}{2}   \\
-2t \cos \frac{k_x}{2} & 0 &   \Gamma^\ast(\textbf{k})   \\
-2t \cos \frac{k_y}{2} & \Gamma(\textbf{k}) &   0   
\end{matrix}
\right)  ,
\end{equation}
with $\Gamma(\textbf{k}) = -4t' \cos \frac{k_x}{2}\cos \frac{k_y}{2} +4i\lambda  \sin \frac{k_x}{2} \sin \frac{k_y}{2} $.

\begin{figure}[htbp]
\center
\includegraphics[width=8.5cm]{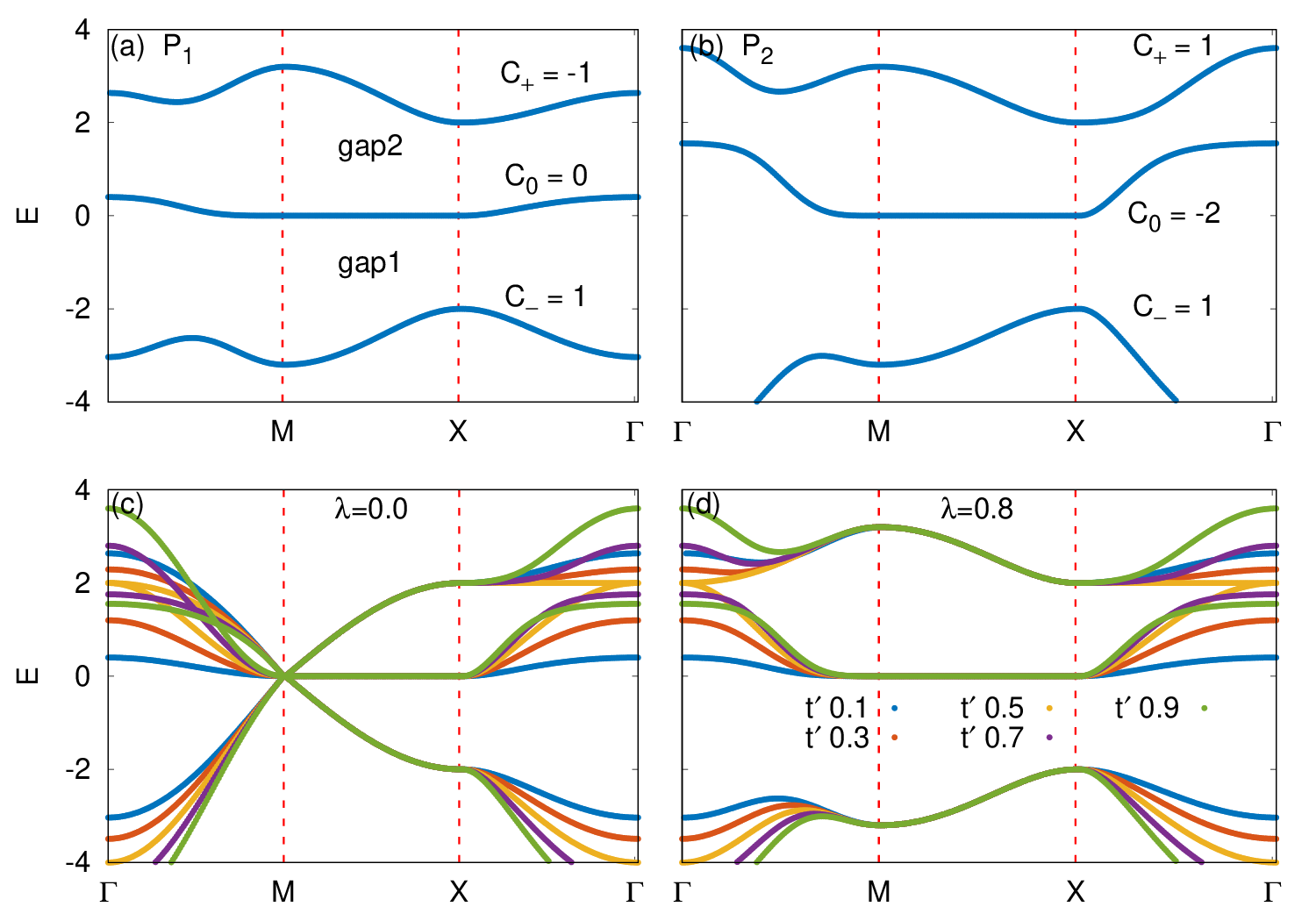}
\caption{Energy Dispersions along the high symmetry lines in the first Brillouin zone. (a) for $P_1$ with $t' = 0.1, \lambda=0.8$. (b) for $P_2$ with $t' = 0.9, \lambda=0.8$.(c) fix parameter $\lambda=0$ and vary $t' = 0.1, 0.3, \cdots, 0.9$. and (d) fix parameter $\lambda=0.8$ and vary $t' = 0.1, 0.3, \cdots, 0.9$. $\mathcal{C}$ denotes the Chern number of corresponding energy branch. $\Gamma(0,0), M(\pi,\pi), X(\pi,0)$ are high symmetry points.}
\label{fig:sp}
\end{figure}

Let us begin with a review of what will happen to the three-band spectrums when $t'$ and $\lambda$ are introduced into the Lieb systems~\cite{Beugeling}. Here, we focus on the parameter space with $t',\lambda \in[0,1]$. Behaviors in a larger parameter space like $t',\lambda \in[-1,1]$ could be readily inferred through symmetric analysis.   First of all, when $t'=0$ and $\lambda \neq 0$, gaps open between upper (lower) dispersive band and the middle flat-band at $M(\pi,\pi)$ point in the first Brillouin zone. These two gaps are nontrivial since the two dispersive bands have their own nonzero Chern number~\cite{Suzuki} and corresponding helical edge states exist. In this case, the bands are well separated,  maintaining particle-hole symmetry, and we denote the Chern number of the lower, middle, and upper bands $\mathcal{C} = (\mathcal{C}_-, \mathcal{C}_0, \mathcal{C}_+)=(1,0,-1)$.
When $t' \neq 0$ and $\lambda = 0$,  the spectrums remain gapless and the middle flat-band develops a maximum at $\Gamma(0,0)$ point(see details in Fig.~\ref{fig:sp}(c)). Increasing $t'$ to $t' = 0.5$, the middle band again touches the upper band  at $\Gamma$. Further enhancing the real NNN hopping $t'$, the intersection point changes from $\Gamma$ to $Q$ which is right on the high symmetry line between the $\Gamma$ and $M$ point.

In the most general case with non-zero $t'$ and $\lambda$, a finite gap(lower gap, denoted as gap1) is always maintained between the lower energy branch and the middle band. This energy gap is independent of $t'$ and grows with the parameter $\lambda$ and reaches saturation -showing no further change-once $\lambda \geq 0.5$. Fig.\ref{fig:sketch-gap}(b) depicts the dependence of another energy gap (upper gap, denoted as gap2, between the upper energy band and the middle branch) on the parameters $\lambda$ and $t'$. Firstly, the dark blue region is gapless since there is no full band gap between the middle and upper band. The absence of a full band gap is evidenced by the fact that the minimum of the upper band is lower in energy than the maximum of the middle band somewhere in the Brillouin zone.
Moreover, two different gapped phases are separated by dashed line with $t' = 0.5$ where gap2 closes  and we select two sets $P_1$ and $P_2$ as reference points. As shown in Fig.\ref{fig:sp}(a)(b), $P_1$ and $P_2$ reside in different topological phases, characterized by distinct Chern numbers for the bands. Clearly, a non-zero ISOC $\lambda$  is essential for opening the gap  whereas $t'$ is primarily responsible for modifying the middle flat band.  Once $\lambda=0$ the gap closes and there is no well defined Chern number for each band, see in Fig.\ref{fig:sp}(c)(d).

\subsection{Construction of transfer-matrix for 2D Lieb lattice with complex NNN hopping}

Based on the transfer-matrix method~\cite{Kramer, JieLiu2020}, we construct the  explicit transfer-matrix $T$ for 2D Lieb lattice with complex NNN hopping.  As usual, we construct a bar shaped system that is a quasi-one-dimensional system of size $ L \times N_y \times N_y$, where the cross section area is $N_y^2$, layer  length $L \gg N_y$. The transfer-matrix  propagates along x direction and according to the Schr\"{o}dinger equation of tight binding equations, the localization length may be calculated from the limiting behavior of products of transfer-matrices for $ L \rightarrow \infty$. In general, it is self-averaging quantity and will not depend on the particular realization of disorders. For a given energy $E$ and disorder strength $W$, the localization length $\xi(E,W)$ is given by $\xi(E,W) = \frac{1}{\gamma_{min}(E,W)}$ with $\gamma_{min}(E,W)$ being the smallest positive Lyapunov exponent. All the Lyapunov  exponent $\gamma_i$ could be obtained by performing QR decomposition on the transfer matrix $T$ with $\gamma_i \simeq \frac{1}{L} \sum_{n=1}^L \ln{\vert R_n \vert_{ii}}$.

The tight-binding model with NN hopping $t$, NNN hopping $t'$ and ISOC $\lambda$ has been presented in Eq.\eqref{eq:tb}. A-A coupling strength is set to be $\delta$.  Substituting  the third equation for $\psi_{m,n}^C$ into the first one of Eq.\eqref{eq:tb}  we obtain:

\begin{eqnarray}
\label{eq:psiA}
\psi_{m+1}^A  &=&  S_{m+1}^{-1} (E-H_m^A) \psi_m^A -S_{m+1}^{-1} S_{m-1}\psi_{m-1}^A  \nonumber \\
&-& S_{m+1}^{-1} R_m \psi_m^B -S_{m+1}^{-1} R_{m-1}\psi_{m-1}^B 
\end{eqnarray}
where we hide the $n$ index to keep the notation concise. Here $S_{m\pm1}$ represents the hopping matrices from slice $A_m$ to slice $A_{m\pm1}$, i.e., $S_{m\pm1}=diag(\delta,\delta,\delta,\dots)$, where $diag$ denotes the diagonal matrix. $R_m$ is the hopping matrix for slice $A_m$ moving forward to slice $B_m$. $H_m^A$ denotes the Hamiltonian of slice $A_m$, 
\begin{widetext}
    \begin{equation}
R_{m } = 
\left(
\begin{array}{cccccc }
g_{m,0} &   \frac{t(t'+i\lambda)}{E-\epsilon_{m,0}^C}  & 0 & 0 & \cdots &  ( \frac{t(t'-i\lambda)}{E-\epsilon_{m,N_y-1}^C} )\\
\frac{t(t'-i\lambda)}{E-\epsilon_{m,0}^C}  & g_{m,1} & \frac{t(t'+i\lambda)}{E-\epsilon_{m,1}^C} & 0  & \cdots  &0\\
0 & \frac{t(t'-i\lambda)}{E-\epsilon_{m,1}^C }& g_{m,2} &\frac{t(t'+i\lambda)}{E-\epsilon_{m,2}^C}& \cdots & 0 \\
\vdots & \vdots& \vdots & \vdots &  \ddots & 0  \\
(\frac{t(t'+i\lambda)}{E-\epsilon_{m,N_y-1}^C} ) & 0 & 0 &  0 &\frac{t(t'-i\lambda)}{E-\epsilon_{m,N_y-2}^C}& g_{m,N_y-1}   \\
\end{array}
\right) 
\end{equation}
\begin{eqnarray}
H_m^A = 
\left(
\begin{array}{cccccc }
f_{m,0} & \delta + \frac{t^2}{E-\epsilon_{m,0}^C} & 0 & 0 & \cdots &  ( \delta + \frac{t^2}{E-\epsilon_{m,N_y-1}^C} )\\
\delta + \frac{t^2}{E-\epsilon_{m,0}^C}  & f_{m,1}  & \delta + \frac{t^2}{E-\epsilon_{m,1}^C}& 0  & \cdots  &0\\
0 & \delta + \frac{t^2}{E-\epsilon_{m,1}^C} & f_{m,2}  &\delta + \frac{t^2}{E-\epsilon_{m,2}^C} & \cdots & 0 \\
\vdots & \vdots& \vdots & \vdots &  \ddots & 0  \\
(\delta + \frac{t^2}{E-\epsilon_{m,N_y-1}^C} ) & 0 & 0 &  0 &\delta + \frac{t^2}{E-\epsilon_{m,N_y-2}^C}  &  f_{m,N_y-1}   \\
\end{array}
\right) 
\end{eqnarray}
\end{widetext}
where the diagonal elements reads $g_{m,n} =  [t(t' - i\lambda)]/(E-\epsilon_{m,n}^C)+[t(t'+i\lambda)]/(E-\epsilon_{m,n-1}^C) -t $, and  $f_{m,n}$ denotes the onsite energy which could be written as $f_{m,n} =  \epsilon_{m,n}^A + t^2/(E-\epsilon_{m,n}^C)  + t^2/(E-\epsilon_{m,n-1}^C)$. The backward hopping matrix from slice A to slice B can be easily obtained, since $R_{m-1}=R_{m}^\ast$, where the symbol $\ast$ denotes the complex conjugate. The above Eq.\eqref{eq:psiA} can be further expressed in matrix form $(\psi_{m+1}^A,\psi_{m}^B,\psi_{m}^A,\psi_{m-1}^B)^T=T_{BA}(\psi_{m}^B,\psi_{m}^A,\psi_{m-1}^B,\psi_{m-1}^A)^T$  and the $T_{BA}$ is the transfer matrix from slice $A_m$ to $B_m$, 
\begin{equation}
T_{BA} = \left(
\begin{array}{cccc}
T_{BA}^{11} &  T_{BA}^{12} & T_{BA}^{13} & T_{BA}^{14}\\
I & 0 & 0 & 0\\
0 & I & 0 & 0\\
0 & 0 & I & 0
\end{array}
\right)
\end{equation}
where $T_{BA}^{11}=- S_{m+1}^{-1} R_m$, $T_{BA}^{12}=S_{m+1}^{-1} (E-H_m^A)$, $T_{BA}^{13}=-S_{m+1}^{-1} R_{m-1}$ and $T_{BA}^{14}=-S_{m+1}^{-1} S_{m-1} $, $I$ is identity matrix. With all the matrix $S_{m\pm1}$, $R_{m}$ and $R_{m-1}$ in hand, $T_{BA} $ has been constructed completely .   

To derive the overall transfer matrix,  the transfer matrix $T_{AB}$ from slice $B_m$ to slice $A_{m+1}$ is also required. Substituting $\psi_{m,n}^C$ into the second equation of Eq.\eqref{eq:tb} in the same manner yields :

\begin{eqnarray}
\label{eq:psiB}
\psi_{m+1}^B &=&  \tilde{S}_{m+1}^{-1} (E-H_m^B) \psi_m^B -\tilde{S}_{m+1}^{-1}\tilde{S}_{m-1}\psi_{m-1}^B \nonumber \\
&-& \tilde{S}_{m+1}^{-1}\tilde{R}_m \psi_m^A - \tilde{S}_{m+1}^{-1}\tilde{R}_{m+1}\psi_{m+1}^A 
\end{eqnarray}
The above equation could also be expressed in matrix form  $(\psi_{m+1}^B,\psi_{m+1}^A,\psi_{m}^B,\psi_{m}^A)^T=T_{AB}(\psi_{m+1}^A,\psi_{m}^B,\psi_{m}^A,\psi_{m-1}^B)^T$  with
\begin{equation}
T_{AB} = \left(
\begin{array}{cccc}
T_{AB}^{11} &  T_{AB}^{12} & T_{AB}^{13} & T_{AB}^{14}\\
I & 0 & 0 & 0\\
0 & I & 0 & 0\\
0 & 0 & I & 0
\end{array}
\right)
\end{equation}
where, $T_{AB}^{11}=- \tilde{S}_{m+1}^{-1}\tilde{R}_{m+1}$, $T_{AB}^{12}=\tilde{S}_{m+1}^{-1} (E-H_B^m)$, $T_{AB}^{13}=-\tilde{S}_{m+1}^{-1} \tilde{R}_{m}$, and $T_{AB}^{14}=-\tilde{S}_{m+1}^{-1} \tilde{S}_{m-1} $.

Similarly, once the Hamiltonian $H_{m}^B$, the hopping matrices  $\tilde{R}_{m}$, $\tilde{R}_{m+1}$ from slice $B$ to slice $A$ and $\tilde{S}_{m+1}$, $\tilde{S}_{m-1}$ are given, the exact transfer matrix $T_{AB}$ can be determined.  By virtue of the Hermitian property of the Hamiltonian in the  2D Lieb lattice, this objective can be readily accomplished, which yields  $H_m^B =Tridiag(L, D, U)= L_m + D_m + U_m$. Here $D_m= diag(\tilde{f}_{m,n})$ is the diagonal onsite Hamiltonian with $(n\in[0,N_y-1])$ and explicitly we have:
\begin{eqnarray}
\tilde{f}_{m,n}&=&\epsilon_{m,n}^B + \frac{{t'}^2+\lambda^2}{E-\epsilon_{m,n}^C}+\frac{{t'}^2+\lambda^2}{E-\epsilon_{m,n-1}^C} \nonumber \\  
&+& \frac{{t'}^2+\lambda^2}{E-\epsilon_{m+1,n}^C}  + \frac{{t'}^2+\lambda^2}{E-\epsilon_{m+1,n-1}^C}  
\end{eqnarray}
$U_{m}$ is the upper diagonal Hamiltonian, under periodic boundary conditions, it can be written as
\begin{eqnarray}
\nonumber
U_{m}=\left(
\begin{array}{ccccccc}
0 & u_{m,0} & 0 & 0 &\cdots&0 &0\\
0 & 0 & u_{m,1} & 0&\cdots&0 &0\\
0 & 0 & 0 & u_{m,n}&\cdots&0 &0\\
\vdots & \vdots & \vdots& \vdots& \ddots& \vdots& \vdots\\
0& 0 & 0 & 0&\cdots &0&u_{m,N_y-2}\\
u_{m,N_y-1}& 0 & 0 & 0&\cdots &0&0	
\end{array} 	
\right)  \\
\end{eqnarray}
where the off-diagonal elements $u_{m,n}$ can be written as 
\begin{eqnarray}
 u_{m,n} = \frac{(t'-i\lambda)^2}{E-\epsilon_{m+1,n}^C} + \frac{(t'+i\lambda)^2}{E-\epsilon_{m,n}^C}  
\end{eqnarray}
Similarly, $L_{m}$ is the lower diagonal  Hamiltonian, under periodic boundary conditions, it can be written as
\begin{eqnarray}
\nonumber
L_{m}=\left(
\begin{array}{ccccccc}
0 & 0 & 0 & 0 &\cdots&0 &l_{m,N_y-1} \\
l_{m,0}  & 0 & 0 & 0&\cdots&0 &0\\
0 & l_{m,1}  & 0 & 0&\cdots&0 &0\\
0 & 0  & l_{m,n} & 0&\cdots&0 &0\\
\vdots & \vdots & \vdots& \vdots& \ddots& \vdots& \vdots\\
0& 0 & 0 & 0&\cdots &l_{m, N_y-2}&0
\end{array}
\right)\\
\end{eqnarray}
where the off-diagonal elements $l_{m,n}$ can be written as 
\begin{eqnarray}
 l_{m,n} = \frac{(t'-i\lambda)^2}{E-\epsilon_{m,n}^C} + \frac{(t'+i\lambda)^2}{E-\epsilon_{m+1,n}^C}  
\end{eqnarray}

The matrices $\tilde{S}_{m+1}$, $\tilde{S}_{m-1}$, $\tilde{R}_{m}$ and $\tilde{R}_{m+1}$ are listed below:
\begin{widetext}
\begin{eqnarray}
\tilde{S}_{m+1} = 
\begin{array}{cccccc}
\left(
\begin{array}{cccccc }
j_{m,0} &  \frac{{t'}^2+\lambda^2}{E-\epsilon_{m+1, 0}^C}  & 0 & 0 & \cdots &  ( \frac{{t'}^2+\lambda^2}{E-\epsilon_{m+1, N_y-1}^C}  )\\
\frac{{t'}^2+\lambda^2}{E-\epsilon_{m+1, 0}^C}  &j_{m,1} & \frac{{t'}^2+\lambda^2}{E-\epsilon_{m+1, 1}^C}  & 0  & \cdots  &0\\
0 & \frac{{t'}^2+\lambda^2}{E-\epsilon_{m+1, 1}^C} & j_{m,2} &\frac{{t'}^2+\lambda^2}{E-\epsilon_{m+1, 2}^C}  & \cdots & 0 \\
\vdots & \vdots& \vdots & \vdots &  \ddots &  \vdots \\
(\frac{{t'}^2+\lambda^2}{E-\epsilon_{m+1, N_y-1}^C}   ) & 0 & 0 &  0 &\frac{{t'}^2+\lambda^2}{E-\epsilon_{m+1, N_y-2}^C} & j_{m,N_y-1}   \\
\end{array}
\right) 
\end{array}
\end{eqnarray}

\begin{eqnarray}
\tilde{S}_{m-1} = 
\begin{array}{cccccc}
\left(
\begin{array}{cccccc }
k_{m,0} &  \frac{{t'}^2+\lambda^2}{E-\epsilon_{m, 0}^C}  & 0 & 0 & \cdots &  ( \frac{{t'}^2+\lambda^2}{E-\epsilon_{m, N_y-1}^C}  )\\
\frac{{t'}^2+\lambda^2}{E-\epsilon_{m, 0}^C}  &k_{m,1} & \frac{{t'}^2+\lambda^2}{E-\epsilon_{m, 1}^C}  & 0  & \cdots  &0\\
0 & \frac{{t'}^2+\lambda^2}{E-\epsilon_{m, 1}^C} & k_{m,2} &\frac{{t'}^2+\lambda^2}{E-\epsilon_{m, 2}^C}  & \cdots & 0 \\
\vdots & \vdots& \vdots & \vdots &  \ddots &  \vdots \\
(\frac{{t'}^2+\lambda^2}{E-\epsilon_{m, N_y-1}^C}   ) & 0 & 0 &  0 &\frac{{t'}^2+\lambda^2}{E-\epsilon_{m, N_y-2}^C} & k_{m,N_y-1}   \\
\end{array}
\right) 
\end{array}
\end{eqnarray}

\begin{eqnarray}
\tilde{R}_{m} = 
\begin{array}{cccccc}
\left(
\begin{array}{cccccc }
v_{m,0} &  \frac{t(t'+i\lambda)}{E-\epsilon_{m, 0}^C}  & 0 & 0 & \cdots &  (  \frac{t(t'-i\lambda)}{E-\epsilon_{m, N_y-1}^C}  )\\
 \frac{t(t'-i\lambda)}{E-\epsilon_{m, 0}^C} & v_{m,1} &  \frac{t(t'+i\lambda)}{E-\epsilon_{m, 1}^C} & 0  & \cdots  &0\\
0 &  \frac{t(t'-i\lambda)}{E-\epsilon_{m, 1}^C} & v_{m,2} & \frac{t(t'+i\lambda)}{E-\epsilon_{m, 2}^C} & \cdots & 0 \\
\vdots & \vdots& \vdots & \vdots &  \ddots &  \vdots  \\
( \frac{t(t'+i\lambda)}{E-\epsilon_{m, N_y-1}^C}    ) & 0 & 0 &  0 & \frac{t(t'-i\lambda)}{E-\epsilon_{m, N_y-2}^C} & v_{m,N_y-1}   \\
\end{array}
\right) 
\end{array}
\end{eqnarray}
\begin{eqnarray}
\tilde{R}_{m+1} = 
\begin{array}{cccccc}
\left(
\begin{array}{cccccc }
w_{m,0} &  \frac{t(t'-i\lambda)}{E-\epsilon_{m+1, 0}^C}  & 0 & 0 & \cdots &  (  \frac{t(t'+i\lambda)}{E-\epsilon_{m+1, N_y-1}^C}  )\\
 \frac{t(t'+i\lambda)}{E-\epsilon_{m+1, 0}^C} & w_{m,1} &   \frac{t(t'-i\lambda)}{E-\epsilon_{m+1, 1}^C} & 0  & \cdots  &0\\
0 &  \frac{t(t'+i\lambda)}{E-\epsilon_{m+1, 1}^C} & w_{m,2} &\frac{t(t'-i\lambda)}{E-\epsilon_{m+1, 2}^C}  & \cdots & 0 \\
\vdots & \vdots& \vdots & \vdots &  \ddots &   \vdots \\
( \frac{t(t'-i\lambda)}{E-\epsilon_{m+1, N_y-1}^C}   ) & 0 & 0 &  0 & \frac{t(t'+i\lambda)}{E-\epsilon_{m+1, N_y-2}^C} & w_{m,N_y-1}   \\
\end{array}
\right) 
\end{array}
\end{eqnarray}
\end{widetext}

where
\begin{eqnarray}
j_{m,n} &=& \frac{(t' -i\lambda)^2}{E-\epsilon_{m+1, n}^C} + \frac{(t' +i\lambda)^2}{E-\epsilon_{m+1, n-1}^C} \nonumber \\
k_{m,n} &=& \frac{(t' -i\lambda)^2}{E-\epsilon_{m, n-1}^C} + \frac{(t' +i\lambda)^2}{E-\epsilon_{m, n}^C} \nonumber\\
v_{m,n} &=& \frac{t(t'-i\lambda)}{E-\epsilon_{m,n-1}^C} + \frac{t(t'+i\lambda)}{E-\epsilon_{m,n}^C} - t \nonumber \\
w_{m,n} &=& \frac{t(t'-i\lambda)}{E-\epsilon_{m+1,n}^C} + \frac{t(t'+i\lambda)}{E-\epsilon_{m+1,n-1}^C}  - t
\end{eqnarray}

At last the total transfer matrix of 2D Lieb lattice can be given by $T=T_{AB}T_{BA}$.

\section{Results}

\subsection{uncorrelated general disorder case}
In most cases, weak diagonal on-site disorders will induce Anderson localization and the flat-band shows different critical behaviors of localization length compared to the dispersive bands~\cite{FlachEPL, FlachPRB}.

To investigate the localization properties of this flat-band system, we firstly introduce uncorrelated onsite real disorders in A, B and C sublattices with $\varepsilon^\sigma_{m,n} \in [-\frac{W}{2},\frac{W}{2}]$, where $\sigma \in \{A,B,C\}$ and $\varepsilon$ are uniform uncorrelated random numbers with disorder strength $W$. The 2D Lieb lattice with $N \times N$ unit cells is in $x-y$ plane with periodic boundary conditions.

\begin{figure}[htbp]
\center
\includegraphics[width=9cm]{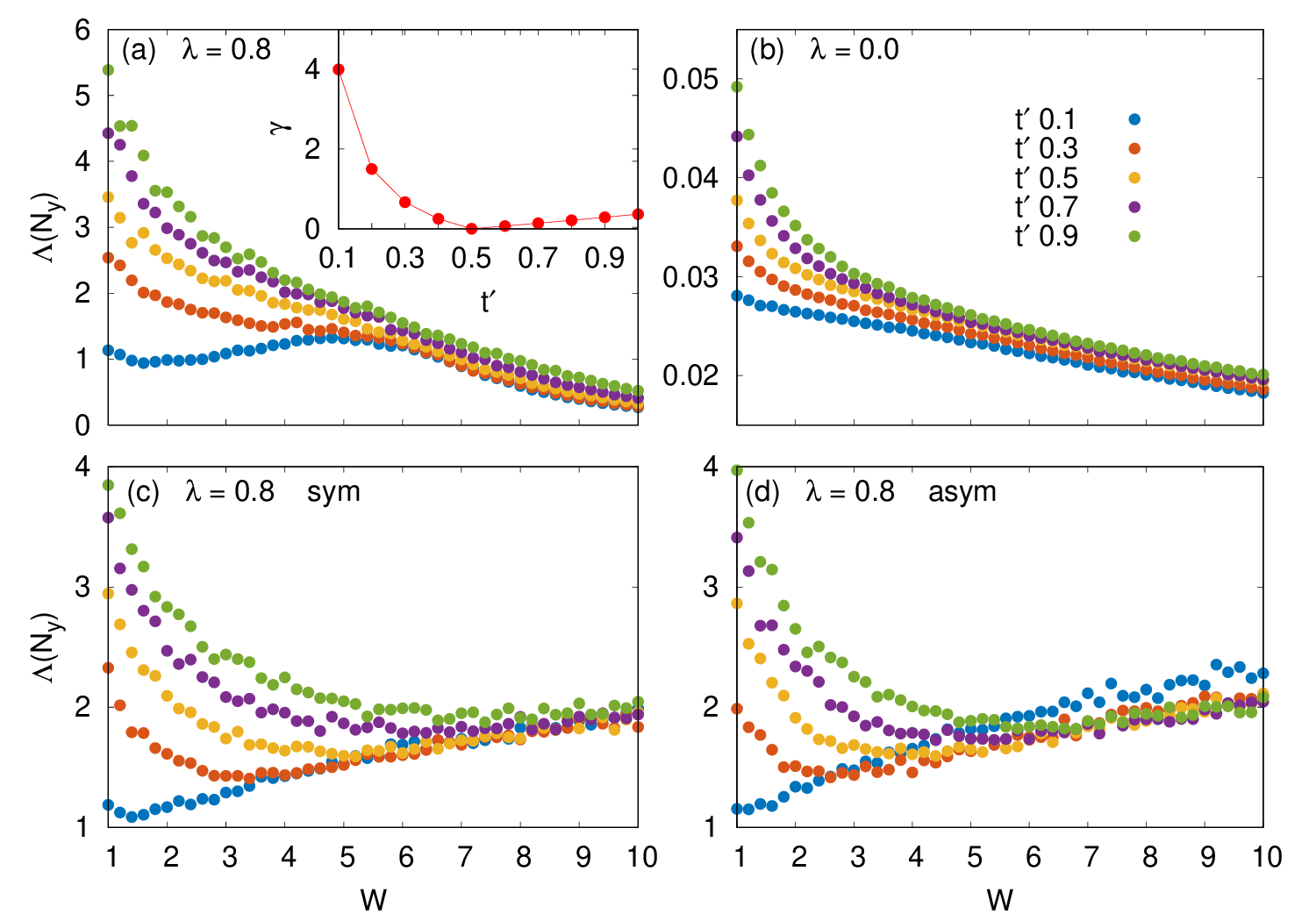}
\caption{Reduced localization length $\Lambda(N_y)$ as a function of disorder strength $W$ with width $N_y = 22$. Panel (a) and (b) depict the results for a fixed parameter $\lambda=0.8$ and $\lambda=0.0$ respectively, with parameter $t'$ varied across the interval $[0.1, 1]$.  The inset shows the flatness ratio $\gamma$ versus  $t'$ with fixed $\lambda = 0.8$.  Panel (c) and (d) depict the results for symmetric and anti-symmetric disorder configurations respectively, i.e., $\varepsilon_{m,n}^B = \pm \varepsilon_{m,n}^C \in [-\frac{W}{2}, \frac{W}{2}], \varepsilon_{m,n}^A=0$. }
\label{fig:localization-length}
\end{figure}

Fig.\ref{fig:localization-length}(a)(b) show the reduced localization length $\Lambda(N_y) = \xi(E,W)/N_y$ with fixed ISOC strength $\lambda=0.8$, $\lambda =0.0$ respectively,  while the NNN real hopping strength $t'$ is varied. Here we focus mainly on the localization behavior of the flat-bands, whose energies are predominantly around $E=0$. 
When $\lambda = 0$, the data overall exhibit a monotonically decreasing trend if we increase the disorder strength $W$. Anderson localization becomes dominant. A detailed inspection of Fig.\ref{fig:sp}(c)  reveals that with the introduction of $t'$, the flat-bands exhibit the least degree of upward bending near the M point when $t'$ is relatively small. This implies that increasing $t'$ tends to suppress localization, as it enhances electron mobility by imparting greater velocity, which in turn destabilizes the flat bands. However, when $\lambda=0.8$ the reduced localization length increased significantly in magnitude. With $\lambda > 0$, at least one band gap opens, giving rise to topological edge states which contribute significantly to transport. In other words, whenever a gap opens, the localization length becomes significantly larger than in the case with $\lambda = 0$. In the region with weaker impurity strengths, several curves with $t' < 0.5$ exhibit a relatively clear separation from the other curves, and there is even a non-monotonic decrease at $t'=0.1$. This separation occurs due to the uneven velocity differences of the flat band near the M point (see details in Fig.\ref{fig:sp}(d)), while the non-monotonic decrease is attributed to the relatively high flatness ratio $\gamma$ of the corresponding flat band, as shown in  Fig.\ref{fig:localization-length}. Here we define flatness-ratio $\gamma = \Delta/W_f$ where $\Delta = \min[gap1, gap2]$ and flat band width $W_f = \max[E_f(k) - E_f(k')]$ with $k,k' \in BZ$,  $E_f$ denotes the middle flat-bands. It is reasonable to expect more robust transport due to large gap and relative narrow band width.

\begin{figure}[htbp]
\center
\includegraphics[width=8.5cm]{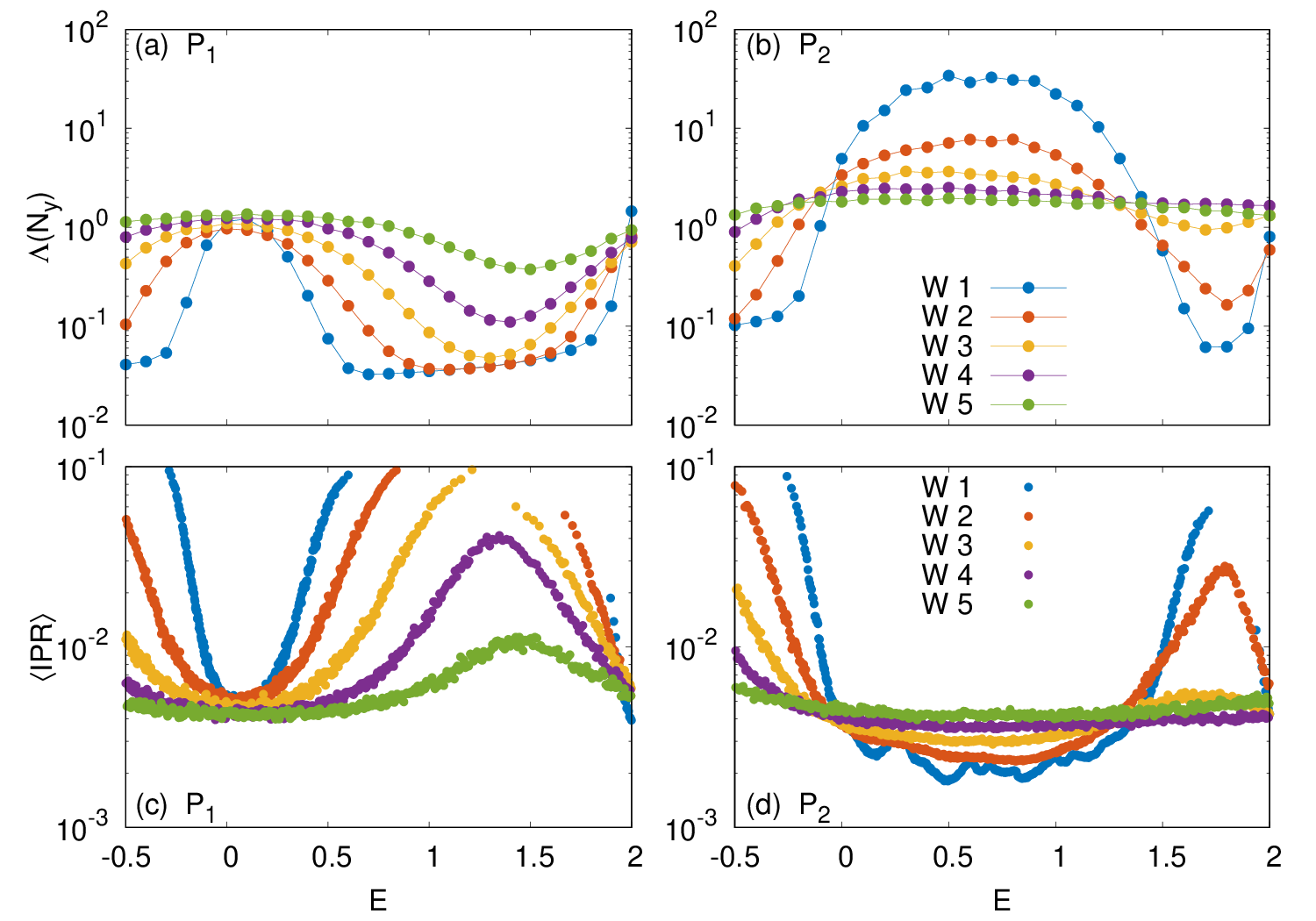}
\caption{(a)(b) Reduced localization length $\Lambda(N_y)$ as a function of energy $E$ for $P_1$ and $P_2$ cases, respectively. Several disorder strengths are considered. Averaged  $\langle \rm{IPR} \rangle$ values for eigenstates with energy $E$ for (c) $P_1$ and (d) $P_2$. }
\label{fig:Lambda-IPR-general}
\end{figure}

Increasing $t'$ will cause the flat-band to bend upward, and the introduction of disorders will also lead to a broadening of the flat band. Therefore, we specifically analyze the localization behavior of the system by calculating $\Lambda(N_y)$ and the inverse participation ratio (IPR) within a certain energy range. The IPR for the $j$th normalized  eigenstate of the disordered system reads: 
\begin{equation}
    \rm{IPR_j} = \sum_{m,n} (\vert \psi_{m,n}^A(j)\vert^4+\vert \psi_{m,n}^B(j)\vert^4+\vert \psi_{m,n}^C(j)\vert^4)
\end{equation}
It is used to analyze the wave-packet distribution and differentiate between localized and extended states.
IPR quantifies the spatial concentration or localization degree of the wave function. For perfectly extended state, the wave function is evenly distributed throughout the system and IPR vanishes in the thermodynamic limit.   
For the localized state, it converges to a finite constant associated with the localization length $\xi$ with  $ \rm{IPR} \propto  \frac{1}{\xi}$.  As demonstrated in Fig.\ref{fig:Lambda-IPR-general}, the global behaviors of reduced localization length $\Lambda(N_y)$ and IPR shows inverse relationships. Compared to $P_2$ case, $P_1$ case exhibits relatively localized behavior, especially for moderate disorder strength $W<4$. The essential reason for this phenomenon can be attributed to the fact that in the $P_1$ case ($t' = 0.1$ is small), the flat band undergoes less distortion, resulting in more pronounced localization effects induced by the flat-band. In addition, $P_1$ case exhibits more robust localization properties, evidenced by the smaller variation of its localization length in magnitude with increasing disorder strength $W$. 

\begin{figure}[htbp]
\center
\includegraphics[width=8.5cm]{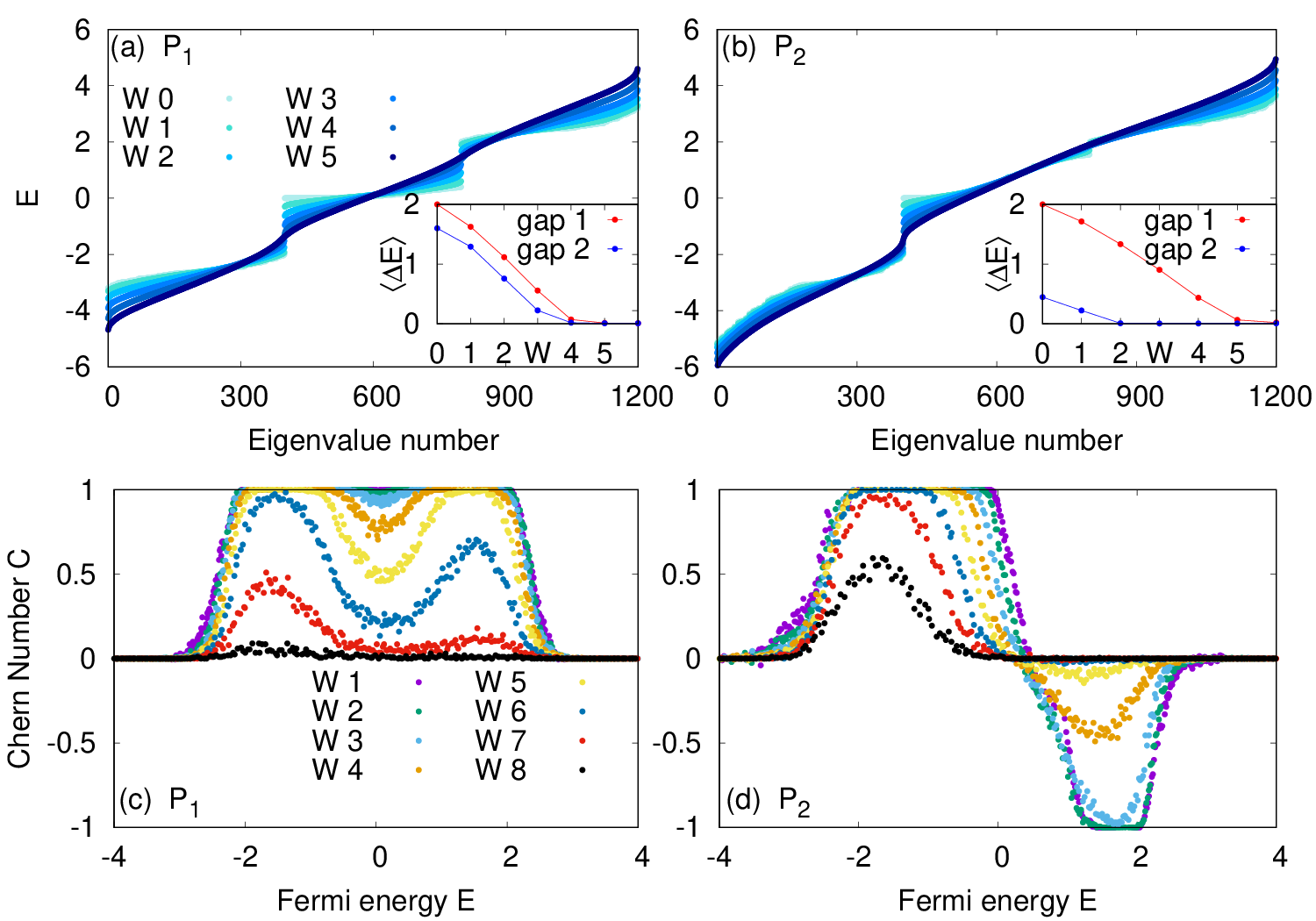}
\caption{The energy spectrum for different disorder strength with $20\times20$ unit cells. (a) for $P_1$ and (b) for $P_2$ case.  The inset depicts the averaged two energy gap  $\langle \Delta E \rangle$ versus disorder strength $W$. (c)(d) Chern number as a function of electron Fermi energy $E$ for two phases under several different disorder strengths. The result is averaged over 200 disorder realizations. }
\label{fig:Chern-number-general}
\end{figure}

To further discuss the topological phase transition, in Fig.\ref{fig:Chern-number-general} we show the energy spectrum of finite Lieb lattice and Chern number.  As is well established, the Chern number serves as a hallmark topological invariant for classifying distinct topological phases in two-dimensional systems. Here we employ coupling-matrix method to calculate the Chern number in real space with disorders~\cite{YiFu}. Consistently, with weak disorder $W<4$ the calculated Chern number of $P_1$ and $P_2$ case are coincide with the results in momentum space $\mathcal{C}_{P_1} = (1,0,-1)$ and $\mathcal{C}_{P_2}=(1,-2,1)$. It should be noted that the Chern number in Fig.\ref{fig:Chern-number-general} is the total Chern number contributed by all occupied bands below the Fermi energy $E$, rather than the Chern number associated with an individual band.
Keep increasing disorder strength, the plateau narrows and finally decreases to zero. The middle energy band spectrum also changes from flat at $W=0$ to dispersive bands with gap closing as the disorder strength increase to a large value. Throughout the entire process, no new non-zero plateaus emerged. 

\subsection{correlated symmetric disorder case}

It has been theoretically predicted and experimentally confirmed that, under certain  correlated disorder, an inverse Anderson transition can occur in a quasi one-dimensional rhombic lattice, supported by Aharonov-Bohm caging~\cite{Longhi,Yanbo}.

The CLS of 2D Lieb model, being eigenstates of the flat band, reside solely on the B and C sublattices with vanishing occupation on the A sublattice . To probe the system, we therefore adopt an disorder configuration that maximally disrupts the flat-band, whereas an alternative  complementary doping scheme  which maintains the flat-band has been reported  in prior work~\cite{JieLiu2022}.
Here we introduce spatially correlated disorder in sublattices B and C with symmetric form, i.e. $\varepsilon_{m,n}^B = \varepsilon_{m,n}^C \in [-\frac{W}{2}, \frac{W}{2}]$ while keep $\varepsilon_{m,n}^A = 0$. The case of anti-symmetric correlated disorder ($\varepsilon_{m,n}^B = -\varepsilon_{m,n}^C$ ) has also been investigated and yields similar results. Therefore, our discussion primarily focuses on the symmetric scenario.

\begin{figure}[htbp]
\center
\includegraphics[width=8.5cm]{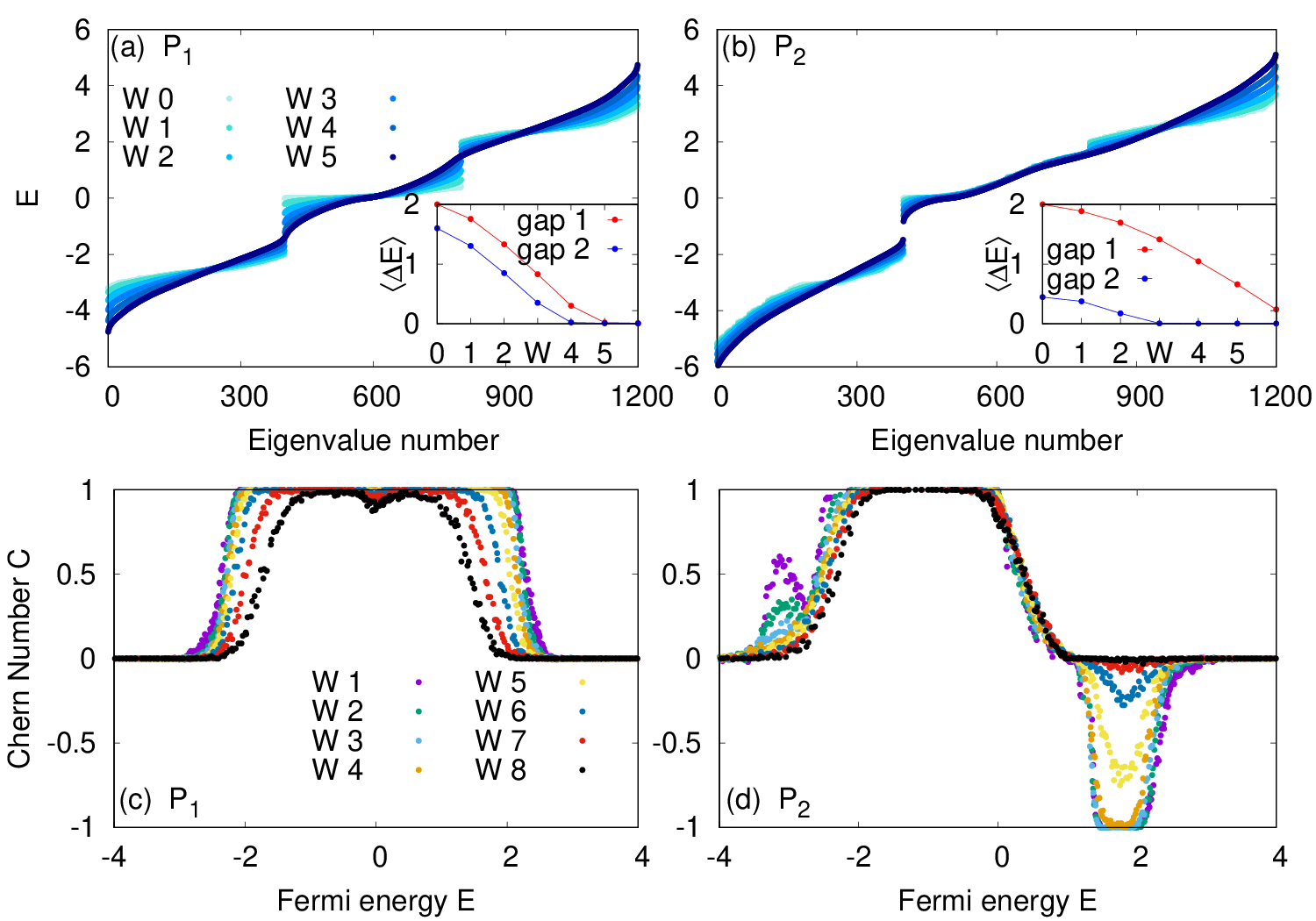}
\caption{The same plots as plotted in Fig.\ref{fig:Chern-number-general} where results come from symmetric disorder configurations, i.e., $\varepsilon_{m,n}^B =  \varepsilon_{m,n}^C \in [-\frac{W}{2}, \frac{W}{2}], \varepsilon_{m,n}^A=0$. }
\label{fig:Chern-number-sym}
\end{figure}

\begin{figure}[htbp]
\center
\includegraphics[width=9cm]{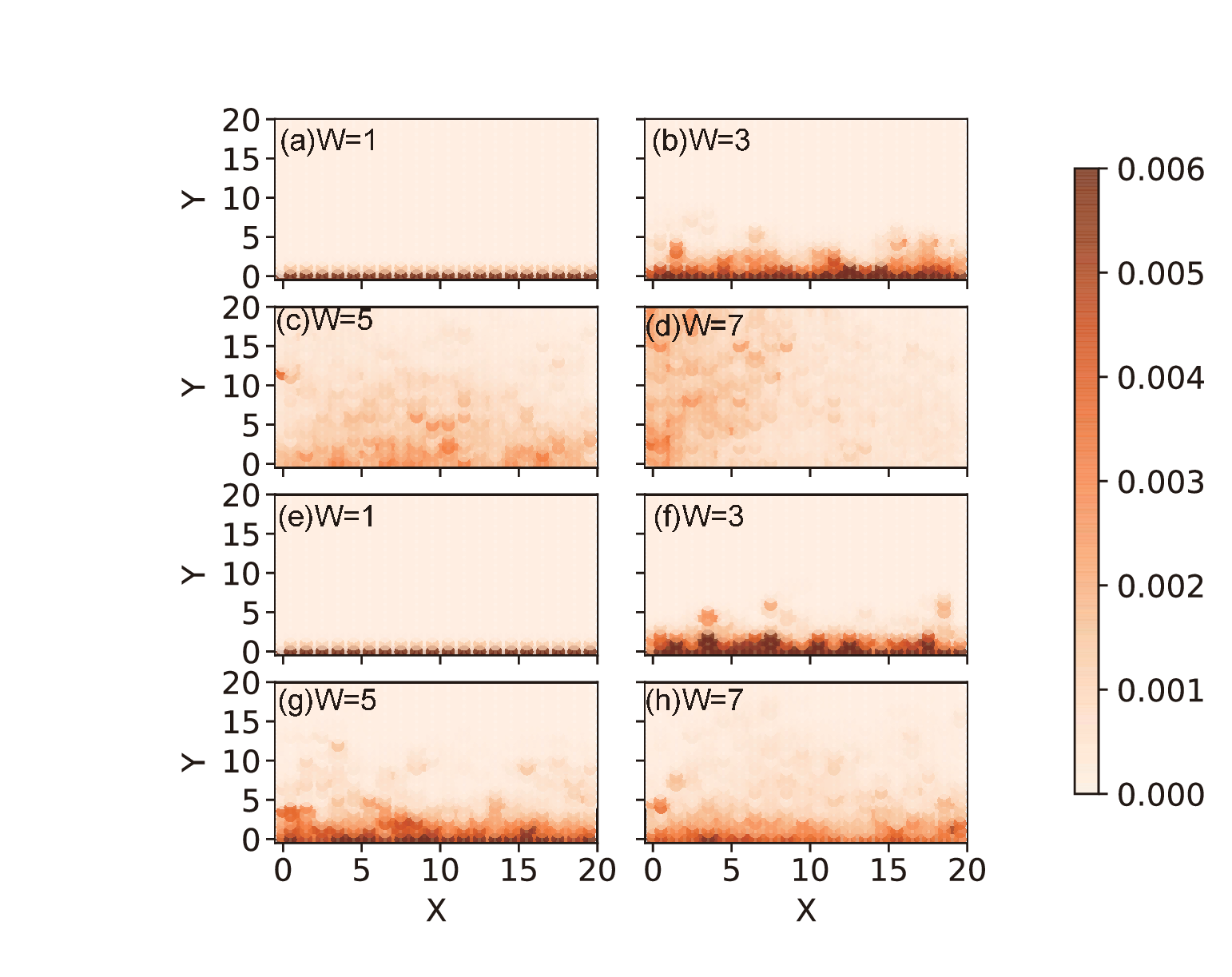}
\caption{One edge state distribution of $P_1$ case in real space under uncorrelated  disorder   with different strength in (a)-(d)  and correlated  symmetric disorder configurations in (e)-(h).}
\label{fig:edge-sym}
\end{figure}

Fig.\ref{fig:localization-length}(c)(d) give the results under correlated disorder configurations. One notable distinction lies in the behavior of the reduced localization length, following an initial decline in the small W region, it plateaus and even begins to rise for $W>5$. In marked contrast, the case of $t'=0.1$ ($P_1$ case) exhibits a persistent upward trend throughout the entire parameter range, indicating the inverse Anderson localization. If we take a closer look at the topological invariants in this case (Fig.\ref{fig:Chern-number-sym}(c)), we find that with increasing the disorder strength, the calculated Chern number $\mathcal{C}=1$ plateau narrows, but remains. This persistent quantized plateau suggests the robustness of the topological phase against disorder and the persistent edge states will facilitates transport.  In addition, the transport is faster due to the larger bandwidth.  Through detailed analysis of the energy spectra, we find the symmetric correlated case exhibits larger band gaps than the uncorrelated case. Moreover,  since gap2 is consistently smaller than gap1,  $\mathcal{C}=1$ plateau is more robust compared to $\mathcal{C}=-1$ in the  $P_2$ case. This robustness difference  is more pronounced for correlated case. 
Specifically, we notice that gap2 closes quickly with increasing disorder strength, whereas gap1 remains open in Fig.\ref{fig:Chern-number-sym}(b). Correspondingly, the computed Chern numbers reflect this contrast: the plateau at 1 remains robust, while the plateau at -1 rapidly decays to 0 as W increases. 
Using the Kwant~\cite{Groth}, we check the edge state distributions in real space with different disorder strengths ($P_1$ case) and verify the robustness of edge state for correlated case in Fig.\ref{fig:edge-sym}.   Clearly, the calculated results are in good agreement with Chern numbers calculation.   These phenomena constitute a stark contrast to the behavior observed in systems with uncorrelated disorder.

\section{Conclusions}
In this study, we construct transfer-matrix for 2D Lieb lattice in the presence of complex NNN coupling and investigate the flat-band localization properties.  When the intrinsic spin-orbit coupling is nonzero ($\lambda \neq 0$), band gaps open along with the emergence of topologically protected edge states, which to some extent counteracts the localization effect in flat bands. Since real NNN hopping $t'$ could modify the flat-band and lift the degeneracy, it introduces velocity  to electrons and further enhances the transport. Under uncorrelated disorder, the localization length reduces with increasing W and finally enters Anderson localized phase. Interestingly, the symmetric/anti-symmetric correlated disorder stabilized the edge state in a way and generate inverse Anderson transition. This result closely corresponds to the discovery of inverse Anderson transition in AB cage~\cite{Longhi,Yanbo} which is driven by anti-symmetric correlated disorder as its the essential condition.  
This study is centered on gapped phase with $\lambda \neq 0$. A double-gap or a single-gap structure (tunning $\lambda$) corresponds to  different numbers of edge states. Interestingly, the localization length calculated at $E=0$ does not exhibit a pronounced variation with the number of edge states. As every real-world system is inevitably subject to dissipation and external perturbations, the investigation of non-Hermitian disorders in flat-band lattices offers a more physically relevant framework. Based on present findings, in future work we will explore the influence of such disorders on flat-band systems and by incorporating dynamical studies, reveal how their localized behaviors differ from those in conventional Hermitian systems. We believe this direction holds promise for contributing meaningfully to the relevant field.

\section{Acknowledgments}
The work is supported by National Natural Science Foundation of China Grant  No. 62588201, the Key Research and Development Program of Guangdong Province (2019B090917007) and the Science and Technology Planning Project of Guangdong Province (2019B090909011). QL was supported by National Natural Science Foundation of China Grant No.12504174. ZXH  was supported by National Natural Science Foundation of China under Grant Nos. 12347101 and 12474140, and the Fundamental Research Funds for the Central Universities under Grant No.
2024CDJXY022.

\end{document}